\documentclass[aps,prd,floatfix,nofootinbib,showpacs,twocolumn,superscriptaddress]{revtex4-1}

\usepackage{mathrsfs}
\usepackage{amsmath}
\usepackage{amssymb}
\usepackage{bbold}
\usepackage{color}
\usepackage{graphicx}
\usepackage{soul}

\usepackage[mathscr,scaled=1.15]{urwchancal}
\DeclareFontFamily{OT1}{pzc}{}
\DeclareFontShape{OT1}{pzc}{m}{it}%
{<-> s * [1.15] pzcmi7t}{}
\DeclareMathAlphabet{\mathpzc}{OT1}{pzc}{m}{it}

\definecolor{purple}{rgb}{0.5,0,0.5}
\definecolor{blue}{rgb}{0.0,0,0.9}
\usepackage[hypertex]{hyperref}
\hypersetup{colorlinks=true, linkcolor=purple, citecolor= purple, urlcolor=blue}

\begin{document}

\title{Structure of the neutral pion and its electromagnetic transition form factor}

\author{Kh{\'e}pani Raya}
\email{khepani@ifm.umich.mx}
\affiliation{Instituto de F\'{\i}sica y Matem\'aticas, Universidad
Michoacana de San Nicol\'as de Hidalgo\\
Edificio C-3, Ciudad Universitaria, C.P. 58040,
Morelia, Michoac\'an, M{\'e}xico}

\author{Lei Chang}
\email{lei.chiong@gmail.com}
\affiliation{School of Physics, Nankai University, Tianjin 300071, China}

\author{Adnan Bashir}
\email{adnan@ifm.umich.mx}
\affiliation{Instituto de F\'{\i}sica y Matem\'aticas, Universidad
Michoacana de San Nicol\'as de Hidalgo\\
Edificio C-3, Ciudad Universitaria, C.P. 58040,
Morelia, Michoac\'an, M{\'e}xico}

\author{J.~Javier Cobos-Martinez}
\email{javiercobos@ifm.umich.mx}
\affiliation{Instituto de F\'{\i}sica y Matem\'aticas, Universidad
Michoacana de San Nicol\'as de Hidalgo\\
Edificio C-3, Ciudad Universitaria, C.P. 58040,
Morelia, Michoac\'an, M{\'e}xico}

\author{L.~Xiomara Guti{\'e}rrez-Guerrero}
\email{laura.gutierrez@unison.mx}
\affiliation{
Departamento de F\'{\i}sica, Universidad de Sonora, Boulevard Luis Encinas J. y Rosales\\
Colonia Centro, Hermosillo, Sonora 83000, M{\'e}xico}

\author{Craig~D.~Roberts}
\email{cdroberts@anl.gov}
\affiliation{Physics Division, Argonne National Laboratory, Argonne
Illinois 60439, USA}

\author{Peter C.~Tandy}
\email{tandy@kent.edu}
\affiliation{
Center for Nuclear Research, Department of Physics, Kent State University, Kent, Ohio 44242, USA}

\date{18 March 2016}

\begin{abstract}
The $\gamma^\ast \gamma \to \pi^0$ transition form factor, $G(Q^2)$, is computed on the entire domain of spacelike momenta using a continuum approach to the two valence-body bound-state problem in relativistic quantum field theory: the result agrees with data obtained by the CELLO, CLEO and Belle Collaborations.  The analysis unifies this prediction with that of the pion's valence-quark parton distribution amplitude (PDA) and elastic electromagnetic form factor, and demonstrates, too, that a fully self-consistent treatment can readily connect a pion PDA that is a broad, concave function at the hadronic scale with the perturbative QCD prediction for the transition form factor in the hard photon limit.   The normalisation of that limit is set by the scale of dynamical chiral symmetry breaking, which is a crucial feature of the Standard Model.  Understanding of the latter will thus remain incomplete until definitive transition form factor data is available on $Q^2>10\,$GeV$^2$.
\end{abstract}


\smallskip


\maketitle

\section{Introduction}
%
%
The neutral pion electromagnetic transition form factor, $G_{\gamma^\ast\gamma \pi^0}(Q^2)$, is a very particular expression of this meson's internal structure.  It is measured in the process $\gamma^\ast_Q \gamma\to \pi^0$, which is fascinating because its complete understanding demands a framework capable of simultaneously combining an explanation of the essentially nonperturbative Abelian anomaly \cite{Adler:1969gk, Bell:1969ts, Adler:2004ihFBS}, which determines $G_{\gamma^\ast\gamma \pi^0}(Q^2\simeq 0)$, with the features of perturbative QCD that govern the behaviour of $G_{\gamma^\ast\gamma \pi^0}(Q^2)$ on the domain of ultraviolet momenta.

In the chiral limit, the Abelian anomaly entails that
\begin{equation}
\label{Ggpg0}
2 f_{\pi}^0 G_{\gamma^\ast\gamma \pi^0}(Q^2=0) = 1\,,
\end{equation}
where $f_\pi^0\approx 0.09\,$GeV is the chiral-limit value of the charged pion's leptonic decay constant; and thereby locks the rate of this transition to the strength of dynamical chiral symmetry breaking (DCSB) in the Standard Model, a phenomenon responsible for the generation of more than 98\% of visible mass \cite{national2012NuclearS, Brodsky:2015aia}.  Corrections to Eq.\,\eqref{Ggpg0}, arising from nonzero and unequal light-quark masses, have been computed: they are small; but, curiously, extant measurements suggest that the calculations actually overestimate their size \cite{Larin:2010kq}.

At the other extreme, the property of factorisation in QCD hard scattering processes leads to an inviolable prediction \cite{Lepage:1980fj}:
\begin{equation}
\label{BLuv}
\exists Q_0>\Lambda_{\rm QCD} \; |\;
Q^2 G(Q^2) \stackrel{Q^2 > Q_0^2}{\approx}  4\pi^2 f_\pi,
\end{equation}
where \cite{Agashe:2014kda} $\Lambda_{\rm QCD} \approx 0.3\,$GeV but the value of $Q_0$ is \emph{a priori} unknown. It is natural to compare Eq.\,\eqref{BLuv} with the analogue for $F_\pi(Q^2)$, the charged-pion elastic electromagnetic form factor \cite{Farrar:1979aw,Lepage:1979zb,Efremov:1979qk,Lepage:1980fj}.  With both normalised to unity at $Q^2=0$, then on any momentum domain for which the asymptotic limit of both is valid, the transition form factor is $\pi/[2\alpha_s(Q^2)]$-times \emph{larger}, where $\alpha_s(Q^2)$ is the QCD running coupling.  At $Q^2=4\,$GeV$^2$, this is a factor of \emph{five}.

The prediction in Eq.\,\eqref{BLuv} and the manner by which it is approached are currently receiving keen scrutiny (\emph{e.g}.\ Refs.\,\cite{Radyushkin:2009zg, Agaev:2010aq, Roberts:2010rnS, Brodsky:2011yv, Bakulev:2011rp, Brodsky:2011xx, Stefanis:2012yw, Bakulev:2012nh, ElBennich:2012ij, Lucha:2013yca, Dorokhov:2013xpa}) following publication of data by the BaBar Collaboration \cite{Aubert:2009mc}.  Whilst those data agree with earlier experiments on their common domain of momentum-transfer \cite{Behrend:1990sr,Gronberg:1997fj}, they are unexpectedly far \emph{above} the prediction in Eq.\,\eqref{BLuv} on $Q^2\gtrsim 10\,$GeV$^2$.  Numerous authors have attempted to reconcile the BaBar measurements with Eq.\,\eqref{BLuv}, typically producing a transition form factor whose magnitude on $Q^2\gtrsim 10\,$GeV$^2$ exceeds the ultraviolet limit, without explaining how that limit might finally be recovered \cite{Radyushkin:2009zg, Agaev:2010aq, Dorokhov:2013xpa} or how their results might be reconciled with modern measurements of $F_\pi(Q^2)$ \cite{Volmer:2000ek, Horn:2006tm, Tadevosyan:2007yd, Huber:2008id}.  Others, however, argue that the BaBar data is not an accurate measure of the transition form factor \cite{Roberts:2010rnS, Brodsky:2011yv, Bakulev:2011rp, Brodsky:2011xx, Stefanis:2012yw, Bakulev:2012nh, ElBennich:2012ij, Lucha:2013yca}.  Significantly, data subsequently published by the Belle Collaboration \cite{Uehara:2012ag} appear to be in general agreement with Eq.\,\eqref{BLuv}.

We reconsider these issues herein using a recently developed method to compute the transition form factor on the entire domain of spacelike momentum transfer \cite{Chang:2013pqS, Cloet:2013ttaS, Chang:2013niaS}.  Consequently, within a single continuum approach to the bound-state problem in relativistic quantum field theory, we arrive at a unified description of $G(Q^2)$ and $F_\pi(Q^2)$, and explain how they connect with these mesons' internal structure through a common pion valence-quark parton distribution amplitude (PDA).


\section{Qualitative features of the transition form factor}
We judge there are sound reasons to expect that the limit in Eq.\,\eqref{BLuv} should either be approached from below or only exceeded marginally, with logarithmic approach to the ultraviolet limit from above in that event.  In order to elucidate, consider first the case in the absence of QCD's scaling violations.  The transition form factor involves only one off-shell photon; and therefore a vector meson dominance (VMD) \emph{Ansatz} for the $Q^2$-dependence of the transition produces
\begin{equation}
\label{GVMDQ}
2 f_\pi G(Q^2)=m_\rho^2/(m_\rho^2 + Q^2)\,.
\end{equation}
Notably, the analogue for $F_\pi(Q^2)$ bounds the empirical data from above; but this is not the case with the transition form factor.

\begin{figure}[t]
\centerline{%
\includegraphics[clip,width=0.47\textwidth]{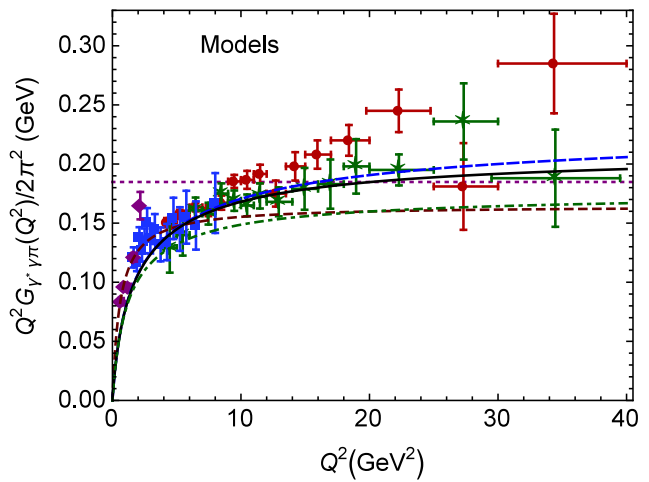}}
\caption{\label{ModelTFF}
Model results for $Q^2 G(Q^2)/(2\pi^2)$ .
Curves:
dotted (purple) -- asymptotic limit, derived from Eq.\,\eqref{BLuv};
dashed (brown) -- VMD result, derived from Eq.\,\eqref{GVMDQ};
dot-dashed (green) -- result obtained from Eq.\,\eqref{anomalytriangle} using a pion Bethe-Salpeter amplitude that generates $\varphi^{\rm cl}(x)$ in Eq.\,\eqref{PDAcl} using Eqs.\,\eqref{pionPDA}, \eqref{piWFA};
long-dashed (blue) -- result obtained from Eq.\,\eqref{anomalytriangle} using a pion PDA that generates $\varphi_\pi(x;\zeta_B)$ in Eq.\,\eqref{phiH} using Eqs.\,\eqref{pionPDA}, \eqref{piWFA};
solid (black) -- result obtained from Eq.\,\eqref{anomalytriangle} using a PDA that evolves from  $\varphi_\pi(x;\zeta_B)$ in Eq.\,\eqref{phiH} to $\varphi_\pi(x;\zeta=Q)$ in Eq.\,\eqref{alphanu} on $Q>\zeta_B$.
Data: BaBar \cite{Aubert:2009mc} -- circles (red); CELLO \cite{Behrend:1990sr} -- diamonds (purple); CLEO \cite{Gronberg:1997fj} -- squares (blue); Belle \cite{Uehara:2012ag} -- stars (green).}
\end{figure}

The VMD form factor in Eq.\,\eqref{GVMDQ} yields a $Q^2\approx 0$ slope for the transition form factor (a neutral pion radius, $r_{\pi^0}$) that is consistent with data \cite{Agashe:2014kda}: $r_{\pi^0} \approx r_{\pi^+}$, so this expression should be a reasonable approximation on $Q^2 \simeq 0$.  However, as evident in Fig.\,\ref{ModelTFF}, it approaches an asymptotic limit of $m_\rho^2/[2f_\pi]$, which is just 90\% of the result associated with Eq.\,\eqref{BLuv}.
Consequently, if $G(Q^2)$ is to approach the limit in Eq.\,\eqref{BLuv} from above in the absence of scaling violations, then it must be influenced by at least three distinct mass scales:
(\emph{i}) an infrared scale that fixes the transition radius, which is smaller than the scale associated with the limit in Eq.\,\eqref{BLuv};
(\emph{ii}) an intermediate scale, marking the point at which nonperturbative aspects of the pion's internal structure begin to take full control of the transition so that the function's fall-off rate may slow and $2f_\pi G(Q^2)$ can thereafter evolve to lie above the ultraviolet limit;
and finally (\emph{iii}) an ultraviolet scale, at which the fall-off rate increases again, as required if $Q^2 G(Q^2)$ is to reach the limit in Eq.\,\eqref{BLuv}.
In the absence of scaling violations, the existence of three mass scales is unlikely, especially since just two scales are evident in predictions \cite{Chang:2013niaS} for $F_\pi(Q^2)$ and the active elements are identical: in both cases, one off-shell photon and the pion wave function.  If three mass scales were possible, they would most probably appear in $F_\pi(Q^2)$ because the associated process is influenced by two pion wave functions in contrast with the single wave function involved in $G(Q^2)$.

Scaling violations are, however, a feature of QCD.  Thus a third scale may appear in connection with the transition form factor, \emph{viz}.\ that associated with the progression to perturbative QCD, which is expressed in the appearance of an additional logarithmic momentum dependence, $[\ln Q^2/\Lambda_{\rm QCD}^2]^{{\mathpzc p}_G}$, that amends dimensional power-law behaviour in the ultraviolet.  The momentum scale for this progression in the charged-pion form factor is $Q^2 \approx 8\,$GeV$^2$ \cite{Chang:2013niaS}.  Universality of pion structure in related processes suggests that a similar scale should be active in the transition form factor.  Considering the neighbourhood $Q^2\simeq 8\,$GeV$^2$, all empirical results for $Q^2 G(Q^2)$ lie below the limit in Eq.\,\eqref{BLuv}; and hence, if logarithmic evolution of $Q^2 G(Q^2)$ becomes established on this domain, then the ultraviolet limit would still be approached slowly from below.  However, if the progression domain is broadened in this case, owing to the presence of just one pion in the transition process, then $Q^2 G(Q^2)$ might grow to marginally exceed $2f_\pi$ before the anomalous dimension, ${\mathpzc p}_G(Q^2)$, finally acquires that asymptotic value which describes a slow approach to the limit in Eq.\,\eqref{BLuv}.

Our picture of $G(Q^2)$ thus involves three mass scales: one associated with the radius, blending effects from the photon-quark interaction and pion structure; another characterising the domain upon which nonperturbative features of pion structure fully control the $\gamma^\ast \gamma \to \pi^0$ transition; and a third typifying the region within which the magnitude of $G(Q^2)$ is still fixed by nonperturbative physics but the momentum dependence of $[Q^2 G(Q^2)-2f_\pi]$ has acquired $[\ln Q^2/\Lambda_{\rm QCD}^2]^{{\mathpzc p}_G}$-damping characteristic of scaling violations in QCD.  Consequently, we expect that the limit in Eq.\,\eqref{BLuv} will either be approached from below or only exceeded slightly, perhaps on a broad domain, with logarithmic approach to the ultraviolet limit in either case.
%


\section{Expressing the transition form factor}
A unification of the $\gamma_Q^\ast \gamma_Q^\ast \to \pi^0$, $\gamma_Q^\ast \gamma \to \pi^0$ transition and charged-pion elastic form factors on $0\leq Q^2 \leq 4\,$GeV$^2$ was accomplished in Ref.\,\cite{Maris:2002mz}.  In fact, the $\gamma_Q^\ast \gamma_Q^\ast \to \pi^0$ transition was computed to arbitrarily large $Q^2$ and shown to approach its QCD hard-photon limit uniformly from below.  However, with the simple algorithm employed therein, it was impossible to extend the $\gamma_Q^\ast \gamma \to \pi^0$ transition and charged-pion elastic form factor calculations into the domain of momenta relevant to modern experiments.  That has now changed following a refinement of known methods \cite{Nakanishi:1963zz,Nakanishi:1969ph,Nakanishi:1971}, described recently in association with a computation of the pion's light-front wave-function \cite{Chang:2013pqS}.  These methods enable simultaneous computation of $G(Q^2)$ and $F_\pi(Q^2)$ to arbitrarily large-$Q^2$, and the correlation of those results with Eq.\,\eqref{BLuv} and the analogous charged-pion formula using the consistently computed distribution amplitude, $\varphi_\pi(x)$.  The latter connection was demonstrated for the charged-pion form factor in Ref.\,\cite{Chang:2013niaS}; and we use that framework herein to complete the picture by calculating $G(Q^2)$.

The $\gamma^\ast \gamma \to \pi^0$ transition form factor is computed from
\begin{equation}
{\mathpzc T}_{\mu\nu}(k_1,k_2) = T_{\mu\nu}(k_1,k_2)+ T_{\nu\mu}(k_2,k_1)\,,
\end{equation}
where the pion's momentum $P=k_1+k_2$, $k_1$ and $k_2$ are the photon momenta, and, at leading order (rainbow-ladder, RL) in the systematic and symmetry-preserving Dyson-Schwinger equation (DSE) truncation scheme reviewed elsewhere \cite{Roberts:2000aa, Chang:2011vu, Bashir:2012fs, Cloet:2013jya},
\begin{align}
\nonumber
T_{\mu\nu}(k_1,k_2) & = \tfrac{e^2}{4\pi^2}\, \epsilon_{\mu\nu\alpha\beta}\,k_{1\alpha} k_{2\beta}\, G(k_1^2,k_1\cdot k_2,k_2^2)\\
\nonumber
& = {\rm tr} \int\frac{d^4 \ell}{(2\pi)^4} \,
i {\cal Q} \chi_\mu(\ell,\ell_1)\, \Gamma_\pi(\ell_1,\ell_2) \, \\
& \quad\quad \times S(\ell_2) \, i {\cal Q} \Gamma_\nu(\ell_2,\ell)\,.
\label{anomalytriangle}
\end{align}
Here $\ell_{1}=\ell+k_1$, $\ell_{2} = \ell - k_2$, ${\cal Q} = {\rm diag}[e_u,e_d] = e\, {\rm diag}[2/3,-1/3]$, where $e$ is the positron charge.
The kinematic constraints are $k_1^2=Q^2$, $k_2^2=0$, $2\, k_1\cdot k_2=- (m_\pi^2+Q^2)$; and the manner by which Eq.\,\eqref{anomalytriangle} provides for a parameter-free realisation of Eq.\,\eqref{Ggpg0} is detailed in Refs.\,\cite{Roberts:1994hh, Maris:1998hc, Holl:2005vu}.  (\emph{N.B}.\ Our calculations are simplified by working with $m_\pi=0$; but this has no material effect on the results.)

The other elements in Eq.\,\eqref{anomalytriangle} are the quark propagator
\begin{equation}
\label{GenSp}
S(p) = -i \gamma\cdot p \, \sigma_V(p^2,\zeta^2)+\sigma_S(p^2,\zeta^2)\,,
\end{equation}
which, consistent with Eq.\,\eqref{anomalytriangle}, is obtained from the rainbow-truncation gap equation ($\zeta$ is the renormalisation scale); the pion Bethe-Salpeter amplitude \cite{Maris:1997hd, Maris:1997tm}
\begin{align}
\nonumber \Gamma_\pi(&\hat k;P)  = i \gamma_5 \left[ E_\pi(\hat k;P) + \gamma\cdot P \, F_\pi(\hat k;P)  \right.\\
& \left. + \gamma\cdot \hat k \hat k\cdot P \, G_\pi(\hat k;P) + \sigma_{\mu\nu} \hat k_\mu P_\nu\, H_\pi(\hat k;P) \right]\,, \label{Gammapi}
\end{align}
computed in RL truncation; and the dressed-quark-photon vertex, $\chi_\mu(k_f,k_i)=S(k_f) \Gamma_\mu(k_f,k_i) S(k_i)$, where $\Gamma_\mu(k_f,k_i)$ is the amputated vertex, which should also be computed in rainbow-ladder truncation.  (\emph{N.B}.\ In Eq.\,\eqref{Gammapi}, $\ell_1=\hat k+\eta P$, $\ell_2=\hat k-(1-\eta P)$, $\eta\in[0,1]$; and Poincar\'e covariance entails that no observable can depend on $\eta$, \emph{i.e}.\ the definition of the relative momentum.)

The leading-order DSE result for the $\gamma^\ast \gamma \to \pi^0$ transition form factor is now determined once an interaction kernel is specified for the rainbow gap equation.  In common with Refs.\,\cite{Chang:2013pqS, Chang:2013niaS}, we employ the kernel explained in Ref.\,\cite{Qin:2011dd} and are therefore spared the need to solve numerically for the dressed-quark propagator and pion Bethe-Salpeter amplitude.  Instead, we can use the perturbation theory integral representations (PTIRs) for $S(p)$ and $\Gamma_\pi(k;P)$ described in Refs.\,\cite{Chang:2013pqS, Chang:2013niaS}, as augmented and employed in Ref.\,\cite{Chang:2013niaS}.  These PTIRs are summarised in Appendix~\ref{appA}.

Such representations are not yet available for the photon-quark vertex; thus we use the following \emph{Ansatz} for the unamputated vertex ($q=k_f-k_i$, $\bar{\mathpzc s}=1-{\mathpzc s}$), expressed completely via the functions which characterise the dressed-quark propagator:
\begin{align}
\nonumber
\chi_\mu(k_f,k_i)   = &  \gamma_\mu \Delta_{k^2 \sigma_V} \\
\nonumber
& + [{\mathpzc s} \, \gamma\cdot k_{f}\gamma_{\mu}\gamma\cdot k_{i}   + \bar{\mathpzc s} \gamma\cdot k_{i}\gamma_{\mu}\gamma\cdot k_{f}]
\Delta_{\sigma_V}  \\
\nonumber
& +
[{\mathpzc s}\,(\gamma\cdot k_f \gamma_\mu + \gamma_\mu \gamma\cdot k_i ) \\
& \quad + \bar{\mathpzc s}\,(\gamma\cdot k_i \gamma_\mu + \gamma_\mu \gamma\cdot k_f )
 ]\, i \Delta_{\sigma_S}\,, \label{ChiAnsatz}
%
%
\end{align}
%
where $\Delta_{F}= [F(k_f^2)-F(k_i^2)]/[k_f^2-k_i^2]$.  A kindred form was useful in computing the charged-pion elastic form factor \cite{Chang:2013niaS}; and our \emph{Ansatz} for $\Gamma_\mu(k_f,k_i)$, Eq.\,(3.84) in Ref.\,\cite{Roberts:1994dr}, is an analogue for the amputated vertex.  Up to transverse pieces associated with ${\mathpzc s}$, $\chi_\mu(k_f,k_i)$ and $S(k_f)\Gamma_\mu(k_f,k_i)S(k_i)$ are equivalent.  Nothing material would be gained herein by making them identical because any difference is power-law suppressed in the ultraviolet; but computational effort would increase substantially.

In using \emph{Ans\"atze} instead of solving the Bethe-Salpeter equation for the vertex, we expedite the computation of $G(Q^2)$.  It is a valid procedure so long as nothing essential to understanding the form factor is lost thereby.  This is established by noting that since the \emph{Ans\"atze} are obtained using the gauge technique \cite{Delbourgo:1977jc}, the vertices derived satisfy the longitudinal Ward-Green-Takahashi identity  \cite{Ward:1950xp, Green:1953te, Takahashi:1957xn}, are free of kinematic singularities, reduce to the bare vertex in the free-field limit, and have the same Poincar\'e transformation properties as the bare vertex.
The \emph{Ans\"atze} have the additional merit that they reproduce the correct ultraviolet limit of the photon-quark interaction when one or both photons is hard (\emph{e.g}.\ Eqs.\,(19) in Ref.\,\cite{Roberts:1998gs} are guaranteed). 
%
A deficiency of these \emph{Ans\"atze} is omission of explicit nonanalytic structures associated with the $\rho$-meson pole.  However, such features only have an impact on $Q^2 r_{\pi^0}^2 \lesssim 1$ and are otherwise immaterial at spacelike momenta \cite{Alkofer:1993gu, Roberts:1994hh, Roberts:2000aa}.  Moreover, salient aspects are included implicitly, \emph{e.g}.\ their influence on pion radii, as explained in connection with Eqs.\,(2.3.39), (2.3.40) in Ref.\,\cite{Roberts:2000aa}.

Owing to the Abelian anomaly \cite{Adler:1969gk, Bell:1969ts, Adler:2004ihFBS}, it is impossible to simultaneously conserve the vector and axial-vector currents associated with Eq.\,\eqref{anomalytriangle}.  We have thus included a momentum redistribution factor in Eq.\,\eqref{ChiAnsatz}:
%
\begin{equation}
{\mathpzc s}=1 +  {\mathpzc s}_0 \exp(-{\mathpzc E}_\pi/M_E)\,,
\end{equation}where ${\mathpzc E}_\pi=Q/2$ is the Breit-frame energy of a massless pion and $M_E=\{p|p^2=M^2(p^2),p>0\}=0.46\,$GeV, with $M(p^2)=\sigma_S(p^2)/\sigma_V(p^2)$, is the computed Euclidean constituent-quark mass \cite{Maris:1997tm}.
Introducing ${\mathpzc s}_0\neq 0$ has an effect equivalent to changing integration variables in Eq.\,\eqref{anomalytriangle}; and hence there is always a value of ${\mathpzc s}_0$ for which the vector currents are conserved and Eq.\,\eqref{Ggpg0} is obtained.  In our case, ${\mathpzc s}_0=1.9$.

It is worth clarifying here that whilst there may appear to be a difference between the photon-quark vertex in Eq.\,\eqref{ChiAnsatz} and that used in Ref.\,\cite{Chang:2013niaS} for the elastic form factor, $F_\pi(Q^2)$, they are actively equivalent.
To this end, we observe that the \emph{Ansatz} in Ref.\,\cite{Chang:2013niaS} is obtained with ${\mathpzc s}=1$ and denote any effect on $F_\pi(Q^2)$ owing to ${\mathpzc s}\neq 1$ in the unamputated photon-quark vertex by $\delta\Lambda$.
The exponential damping factor in $\delta\Lambda$ guarantees that ${\mathpzc s}\approx 1$ for $Q^2 > 4\,$GeV$^2$. Therefore, it is only necessary to consider a possible impact of $\delta\Lambda$ on $F_\pi(Q^2)$ within $0\leq Q^2 \leq 4\,$GeV$^2$.
Here, since $\delta\Lambda$ is purely transverse, it cannot affect the behavior of $F_\pi(Q^2)$ at $Q^2 = 0$ owing to the vector Ward-Green-Takahashi identity.
In considering the intermediate $Q^2$ domain, recall that the pion's Bethe-Salpeter amplitude involves four terms $E_\pi$, $F_\pi$, $G_\pi$, $H_\pi$, Eq.\,\eqref{Gammapi}; and a little spinor algebra shows that $\delta\Lambda$ only contributes to the quadratic $F_\pi \times F_\pi$, $G_\pi \times G_\pi$ terms in the expression for $F_\pi(Q^2)$.
In quantities that involve two pion Bethe-Salpeter amplitudes, such as $F_\pi(Q^2)$, the $E_\pi \times E_\pi$ term dominates at low-$Q^2$.  There are small contributions from the cross-terms $E_\pi \times F_\pi$, $E_\pi \times G_\pi$; but the quadratic terms $F_\pi \times F_\pi$, $G_\pi \times G_\pi$ are negligible, as are all terms involving $H_\pi$.
Finally, it is known that the impact of any term involving $F_\pi$, $G_\pi$ is only felt in the elastic form factor on $Q^2>8\,$GeV$^2$ \cite{Maris:1998hc}, which is far beyond the upper bound of influence allowed by the exponential damping factor.
This reasoning establishes that ${\mathpzc s}\neq 1$ cannot have any material impact on the elastic pion form factor, and hence Eq.\,\eqref{ChiAnsatz} is practically equivalent to the vertex  employed in Ref.\,\cite{Chang:2013niaS}.\footnote{
On the other hand, if one reverses the perspective, \emph{i.e}.\ uses precisely the vertex in Ref.\,\cite{Chang:2013niaS} to compute the transition form factor, then the calculated result for $Q^2 G(Q^2)$ underestimates extant data on $Q^2\in [0,4]\,$GeV$^2$; but nothing is otherwise altered, including all aspects of the discussion herein that concern the approach to the limit in Eq.\,\eqref{BLuv}.}

With all elements in Eq.\,\eqref{anomalytriangle} expressed via a generalised spectral representation, computation of $G(Q^2)$ reduces to the task of summing a series of terms, all of which involve a single four-momentum integral.  The integrand denominator in every term is a product of $\ell$-quadratic forms, each raised to some power.  Within each such term, one uses a Feynman parametrisation in order to combine the denominators into a single quadratic form, raised to the appropriate power.  A suitably chosen change of variables then enables straightforward evaluation of the four-momentum integration using standard algebraic methods.  After calculation of the four-momentum integration, evaluation of the individual term is complete after one computes a finite number of simple integrals; namely, the integrations over Feynman parameters and the spectral integral.  The complete result for $G(Q^2)$ follows after summing the series.


\section{Evolution of the pion Bethe-Salpeter amplitude}
The pion's valence-quark twist-two PDA can be expressed \cite{Chang:2013pqS}:
\begin{align}
\nonumber
 f_\pi \varphi_\pi(x;\zeta)  & = N_c {\rm tr}\,
Z_2 \! \int_{dk}^\Lambda \!\!\delta_n^{x}(k_\eta) \,\gamma_5\gamma\cdot n\, \\
&  
\quad \quad \times S(k_\eta) \Gamma_\pi(\hat k;P) S(k_{\bar \eta})\,,
\label{pionPDA}
\end{align}
where $N_c=3$; the trace is over spinor indices;
$Z_{2}(\zeta,\Lambda)$ is the quark wave-function renormalisation constant;
$\int_{dk}^\Lambda$ is a Poincar\'e-invariant regularisation of the four-dimensional integral, with $\Lambda$ the ultraviolet regularisation mass-scale;
$\delta_n^{x}(k_\eta):= \delta(n\cdot k_\eta - x \,n\cdot P)$, with $n^2=0$, $n\cdot P = -m_\pi$;
and $\hat k = [k_\eta+k_{\bar\eta}]/2$, $k_\eta = k + \eta P$, $k_{\bar\eta} = k - (1-\eta) P$.

The evolution of the PDA with the resolving scale $\zeta$ is explained in Refs.\,\cite{Lepage:1979zb, Efremov:1979qk, Lepage:1980fj}.  It is logarithmic; and whilst Poincar\'e covariant computations using a renormalisation-group-improved RL truncation produce the same matrix-element power-laws as perturbative QCD, they fail to reproduce the full anomalous dimensions \cite{Lepage:1980fj}.  Typically \cite{Maris:1998hc, Chang:2013pqS, Chang:2013niaS}, the RL approximation to a matrix element underestimates the rate of its logarithmic flow with the active momentum scale because the approximation omits gluon-splitting diagrams.

Owing to Eq.\,\eqref{pionPDA}, the pion's Poincar\'e covariant Bethe-Salpeter wave function must evolve with $\zeta$ in the same way as $\varphi_\pi$; but that constraint has hitherto been overlooked in computations of observables using continuum methods in QCD.  Evolution enables the dressed-quark and -antiquark degrees-of-freedom, in terms of which the wave function is expressed at a given scale $\zeta^2=Q^2$, to split into less well-dressed partons via the addition of gluons and sea quarks in the manner prescribed by QCD dynamics.  Such effects are incorporated naturally in bound-state problems when the complete quark-antiquark scattering kernel is used; but aspects are lost when that kernel is truncated, and so it is with the RL truncation.

The impact of this realisation on $G(Q^2)$ is readily illustrated.  Consider the Bethe-Salpeter wave function constructed with
{\allowdisplaybreaks
\begin{subequations}
\label{piWFA}
\begin{align}
S(k) & = 1/[i\gamma\cdot k + M] \,,\\
\label{pionBSAmodel}
\Gamma_\pi(k;P) & = i\gamma_5\,\frac{M}{{\mathpzc n}_\pi\, f_\pi}\,\int_{-1}^1\! dz\,\rho_\nu(z)\frac{M^2}{(k + z P/2)^2+\Lambda_\pi^2}\,,\\
\rho_\nu(z) & = \frac{\Gamma(\tfrac{3}{2}+\nu)}{\sqrt{\pi}\,\Gamma(1+\nu)}\,(1-z^2)^\nu\,, \label{rhonu}
\end{align}
\end{subequations}}
\hspace*{-0.5\parindent}where ${\mathpzc n}_\pi$ is the canonical normalisation constant \cite{LlewellynSmith:1969az}.  Inserting these formulae into Eq.\,\eqref{pionPDA}, then with $\nu=1$ the result is the PDA associated with QCD's conformal limit  \cite{Chang:2013pqS}, \emph{viz}.\
\begin{equation}
\label{PDAcl}
 \varphi_\pi(x)=\varphi^{\rm cl}(x)=6x(1-x)\,.
 \end{equation}
 It is therefore worthwhile to compute the transition form factor in Eq.\,\eqref{anomalytriangle} using Eqs.\,\eqref{piWFA} with $\nu=1$, which corresponds to a conformal-limit pion Bethe-Salpeter wave function, \emph{i.e}.\ a wave function that is frozen to produce Eq.\,\eqref{PDAcl} at all scales $\zeta$.
 The result yields the dot-dashed (green) curve in Fig.\,\ref{ModelTFF}, which is a monotonically increasing, concave function that approaches the asymptotic limit associated with Eq.\,\eqref{BLuv} from below.

In contrast, the PDA at any scale realisable with existing facilities is very different from $\varphi^{\rm cl}(x)$.  In fact, at $\zeta_B\simeq 2\,$GeV the pion PDA takes the form \cite{Mikhailov:1986be, Petrov:1998kgS, Braun:2006dg, Brodsky:2006uqa, Chang:2013pqS, Cloet:2013ttaS}
\begin{equation}
\label{phiH}
\varphi_\pi(x;\zeta_B) \approx \tfrac{8}{\pi} \, \sqrt{x(1-x)}\,.
\end{equation}
This PDA is obtained from Eq.\,\eqref{pionPDA} by using \mbox{$\nu=-\tfrac{1}{2}$} in Eqs.\,\eqref{piWFA}.  Moreover, with a dressed mass $M=0.4\,$GeV, a scale typical of DCSB, and $P^2=-m_\pi^2$, $m_\pi=0.14\,$GeV, then $\Lambda_\pi=0.52\,M$ yields $f_\pi=92\,$MeV.  Eqs.\,\eqref{piWFA} can now be used to compute $G(Q^2)$ via Eq.\,\eqref{anomalytriangle} using a Bethe-Salpeter wave function that is frozen to produce Eq.\,\eqref{phiH} at all scales $\zeta$.  The result is the long-dashed (blue) curve in Fig.\,\ref{ModelTFF}.  Like the prediction obtained using $\nu=1$, this curve is monotonically increasing and concave, and approaches its asymptotic limit from below.  The difference is that the asymptotic limit is not that associated with Eq.\,\eqref{BLuv}. Instead, this curve approaches $(8/3) f_\pi$ as $Q^2\to \infty$.

As a last illustration, consider $2\rho_\nu(z)=\delta(1+z)+\delta(1-z)$ in Eq.\,\eqref{pionBSAmodel}.  This gives $\varphi_\pi(x)=1$ via Eq.\,\eqref{pionPDA}, which is also the result obtained using a translationally invariant regularisation of a momentum-independent quark-quark interaction \cite{Roberts:2010rnS}.
With this input, Eq.\,\eqref{anomalytriangle} yields $Q^2 G(Q^2)  \propto [\ln Q^2/M^2]^2$, and the mass-scale $M$  can be tuned to reproduce the BaBar data in Fig.\,\ref{ModelTFF}.  Notwithstanding that, the complete curve is also monotonically increasing and concave, and approaches its asymptotic limit from below \cite{Roberts:2010rnS}.  Notably, the BaBar data have often been used to justify a ``flat-top'' pion PDA: $\varphi_\pi(x)\approx 1$, \emph{e.g}.\ Refs.\,\cite{Radyushkin:2009zg, Agaev:2010aq, Dorokhov:2013xpa}.  Employing the factorised hard-scattering formula in this case, one finds $Q^2 G(Q^2)  \propto [\ln Q^2/M^2]$, a result which highlights an observation made above, \emph{viz}.\ Poincar\'e covariant treatments of the triangle diagram expressed by Eq.\,\eqref{anomalytriangle} typically yield the correct power-law but produce an erroneous value of the anomalous dimension.

Evidently, any computation of $G(Q^2)$ via Eq.\,\eqref{anomalytriangle} which uses a Bethe-Salpeter amplitude that does not evolve with the resolution scale, $\zeta^2=Q^2$, produces a curve $Q^2 G(Q^2)$ which approaches its asymptotic limit from below; but the value of that limit depends on the model used for the pion's (frozen) Bethe-Salpeter wave function \cite{Roberts:2010rnS}.

Evolution of the interaction current and vertices in Eq.\,\eqref{anomalytriangle} is missing from such model calculations.  The impact of the associated dynamics may be incorporated by recalling that the PDA at a new scale $\zeta>\zeta_B$ can be obtained from $\varphi_\pi(x;\zeta_B)$ by using the ERBL evolution equations \cite{Lepage:1979zb,Efremov:1979qk,Lepage:1980fj}.  In the context of Eqs.\,\eqref{pionPDA}, \eqref{piWFA} this can be translated into an evolution of $\nu$ in Eq.\,\eqref{rhonu}, \emph{i.e}.\ one can reproduce any concave PDA $\varphi_\pi(x;\zeta)$,  obtained via ERBL evolution of $\varphi_\pi(x;\zeta_B)$, by using a suitably chosen value of $\nu(\zeta)$.
Solving for $\nu(\zeta)$ is straightforward because at any $\zeta>\zeta_B$, the ERBL-evolved form of $\varphi_\pi(x;\zeta_B)$ is
\begin{equation}
\varphi_\pi(x;\zeta) =  [x(1-x)]^{\alpha(\zeta)}/B(1+\alpha(\zeta),1+\alpha(\zeta))\,.
\label{alphanu}
\end{equation}
One may readily compute the trajectory $\alpha(\zeta)$.  Equally, for any $\nu$ in Eqs.\,\eqref{piWFA}, one can determine $\alpha(\nu)$: $\nu(\zeta)$ is then that number which produces $\alpha(\zeta)$.  On the domain $\alpha(\zeta) \in (0.3,0.8)$ that is relevant herein, we find
\begin{equation}
\label{nuofalpha}
\nu(\alpha(\zeta)) = - [ 5.4 - 6.6\,\alpha(\zeta) ]/[ 5.9-2.8 \,\alpha(\zeta)]\,.
\end{equation}
%

Employing Eqs.\,\eqref{piWFA}, \eqref{nuofalpha} in Eq.\,\eqref{anomalytriangle}, we obtain a result for $G(Q^2)$ that expresses the impact of a Bethe-Salpeter wave function which evolves and thereby interpolates between $\varphi_\pi(x;\zeta_B)$ in Eq.\,\eqref{phiH} and $\varphi^{\rm cl}(x)$ in Eq.\,\eqref{PDAcl}.  This is the solid (black) curve in Fig.\,\ref{ModelTFF}.  Specifically, that curve is obtained by using $\nu=-1/2$ to define the pion's Bethe-Salpeter wave function on $Q=\sqrt{Q^2}<\zeta_B$, whilst thereafter the wave function is defined using $\nu(\zeta=Q)$ in Eqs.\,\eqref{piWFA}.  Obtained this way, $Q^2 G(Q^2)$ is monotonically increasing and concave on the domain depicted, and reaches a little above the asymptotic limit associated with Eq.\,\eqref{BLuv}.  The growth is logarithmically slow, however; and whilst the curve remains a line-width above the asymptotic limit on a large domain, logarithmic growth eventually becomes suppression, and the curve thereafter proceeds towards the QCD asymptotic limit from above.

Now consider these remarks in the context provided by the leading-twist expression for the transition form factor \cite{Lepage:1980fj}:
\begin{align}
\label{HardG}
G(Q^2) = 4\pi^2f_\pi\int_0^1 \! dx \,T_H(x,Q^2,\alpha(\zeta);\zeta)\,\varphi_\pi(x;\zeta)\,,
\end{align}
where $T_H(\zeta)$ is the photon$+$quark$+$antiquark scattering amplitude appropriate to the scale $\zeta$.  On the domain $\Lambda_{\rm QCD}/\zeta \simeq 0$, $T_H(\zeta) = (e_u^2-e_d^2)/(xQ^2)$.  However, this is far from an accurate representation of the scattering amplitude at an hadronic scale, $\zeta=\zeta_B \approx 2\,$GeV, a fact made plain by Eq.\,\eqref{anomalytriangle}, which involves nonperturbatively dressed propagators and vertices.  Indeed, as elucidated in analyses of the chiral condensate \cite{Brodsky:2010xf,Brodsky:2012ku}, from a light-front perspective this dressing corresponds to inclusion of infinitely many Fock-space components in the description of the pion bound-state, its constituents and their interactions.

Owing to convergence issues connected with the need to extrapolate from $\Lambda_{\rm QCD}/\zeta \simeq 0 \to \Lambda_{\rm QCD}/\zeta_B$, a twist expansion cannot systematically connect Eq.\,\eqref{HardG} with Eq.\,\eqref{anomalytriangle}.  On the other hand, if RL-truncation DSE solutions are used for the dressed propagators and vertices in Eq.\,\eqref{anomalytriangle}, then one arrives at Eq.\,\eqref{HardG} on $\Lambda_{\rm QCD}/\zeta \simeq 0$, except for a mismatch $\sim [\ln Q^2/\Lambda_{\rm QCD}^2]^{{\mathpzc p}_G}$.  This discrepancy originates in the failure of RL truncation to reproduce the complete array of gluon and quark splitting effects contained in QCD and hence its failure to fully express interferences between the anomalous dimensions of those $n$-point Schwinger functions which are relevant in the computation of a given scattering amplitude.  It is ameliorated by the procedure discussed in connection with Eq.\,\eqref{alphanu}.  Importantly, if a similar procedure is employed in revisiting the kindred calculation of the pion form factor \cite{Chang:2013niaS}, one can also correct the $\ln Q^2$-evolution of the RL prediction for $F_\pi(Q^2)$.


\section{Numerical results and discussion}
\begin{figure}[t]
\centerline{%
\includegraphics[clip,width=0.47\textwidth]{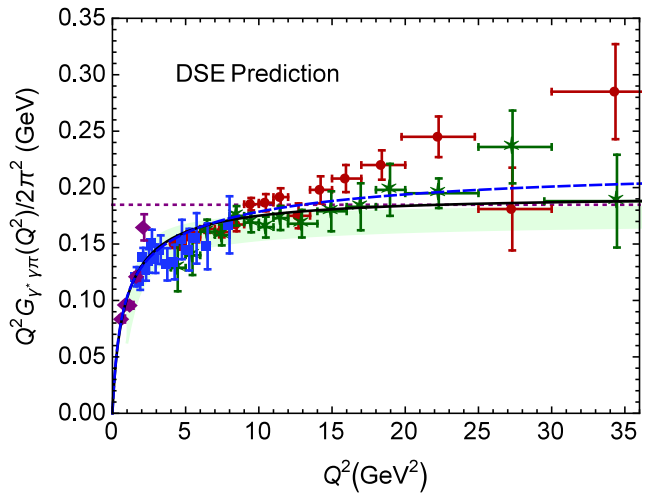}}
\caption{\label{figDSEprediction}
Prediction for $Q^2 G(Q^2)/(2\pi^2)$ .
Curves:
solid (black) -- result obtained from Eq.\,\eqref{anomalytriangle} using RL-truncation propagators, amplitudes and vertices, and ERBL evolution of the pion Bethe-Salpeter amplitude;
long-dashed (blue) -- result obtained without that evolution;
dotted (purple) -- asymptotic limit, derived from Eq.\,\eqref{BLuv}.
Data: BaBar \cite{Aubert:2009mc} -- circles (red); CELLO \cite{Behrend:1990sr} -- diamonds (purple); CLEO \cite{Gronberg:1997fj} -- squares (blue); Belle \cite{Uehara:2012ag} -- stars (green).
The shaded (green) band is described in Ref.\,\cite{Bakulev:2012nh}.
}
\end{figure}

We have calculated the $\gamma^\ast \gamma\to \pi^0$ transition form factor using Eq.\,\eqref{anomalytriangle} with the propagators, amplitudes, vertices, and PTIR-based computational scheme detailed in Sec.\,3.  In this RL analysis, the pion Bethe-Salpeter amplitude yields a PDA via Eq.\,\eqref{pionPDA} that is described by Eq.\,\eqref{alphanu} with $\alpha(\zeta_B)=0.3$ \cite{Chang:2013pqS}.  The result is the dashed (blue) curve depicted in Fig.\,\ref{figDSEprediction}: $G(0)=0.50$ and $r_{\pi^0} =0.68\,$fm, which is practically identical to $r_{\pi^+}$ computed in the same scheme \cite{Chang:2013niaS}.  (\emph{N.B}.\ Our framework enables computation of $G(Q^2)$ $\forall Q^2\geq 0$ and readily admits continuation to timelike momenta.)

The solid (black) curve in Fig.\,\ref{figDSEprediction} is obtained from the dashed curve via multiplication by a $Q^2$-dependent evolution factor.  That factor is a ratio: the denominator is $G(Q^2)$ computed using Eqs.\,\eqref{anomalytriangle}, \eqref{piWFA} with a frozen value of $\alpha=0.3$; and the numerator is $G(Q^2)$ calculated using those same equations but with $\nu$ determined via Eq.\,\eqref{nuofalpha} as $\alpha$ experiences one-loop ERBL evolution on $Q^2>\zeta_B^2$.  This solid curve expresses our final prediction for the transition form factor.  
(\emph{N.B}.\ A ready and fair estimate of the impact of this evolution is obtained by multiplying the result for $G(Q^2)$ obtained with a frozen PDA by an average of the ERBL evolution profiles for the leading, relevant nontrivial moments of $\varphi_\pi(x;\zeta=Q)$, \emph{viz}.\ $\langle 1/x\rangle$ and $\langle (2x-1)^2\rangle$ .)

Our prediction for the transition form factor behaves in precisely the manner one would expect based on the analyses of simple models described in connection with Fig.\,\ref{ModelTFF}.  Moreover, notwithstanding the fact that $G(Q^2)$ was evaluated using a framework which produces a pion PDA that is a broad, concave function at the hadronic scale $\zeta_B$: the calculated transition form factor does not materially exceed the asymptotic limit in Eq.\,\eqref{BLuv}; and this same approach explains both existing measurements of $F_\pi(Q^2)$ and its hard-photon limit.

In the context of a survey of theoretical analyses of the $\gamma^\ast \gamma \to \pi^0$ transition \cite{Bakulev:2012nh}, our prediction is a member of that class of studies, denoted by the (green) shaded band  which are consistent with all non-BaBar data and confirm the standard QCD factorisation result in Eq.\,\eqref{BLuv}.  In particular, the solid (black) curve in Fig.\,\ref{figDSEprediction} is similar to the light-cone sum rules result of Ref.\,\cite{Bakulev:2011rp} on their common domain: sum rules analyses are restricted to  $Q^2 \gtrsim 1\,$GeV$^2$.  In this connection, it is also worth reiterating DSE predictions for some of the moments
$\langle \xi^n \rangle = \int_0^1du\, (2u-1)^n \varphi_\pi(u;\zeta_B)$,
\emph{viz}.\
\begin{equation}
\label{moments}
\begin{array}{l|ccc}
    & \langle \xi^2\rangle & \langle \xi^4\rangle & \langle \xi^6\rangle \\\hline
{\rm RL} & 0.28 & 0.15 & 0.098\\
{\rm DB} & 0.25 & 0.13 & 0.078\\
\end{array}\,,
\end{equation}
where ``DB'' indicates results obtained using the form of $\varphi_\pi(u;\zeta_B)$ in Eq.\,\eqref{phiH}, which is the PDA produced by the most sophisticated DCSB-improved DSE kernels that are currently available \cite{Chang:2013pqS}.\footnote{The function in Eq.\,\eqref{phiH} is not precisely that presented in Eq.\,(15) of Ref.\,\cite{Chang:2013pqS}.  However, as illustrated by the right panel of Fig.\,5.3 in Ref.\,\cite{Cloet:2013jya}, the two forms are not realistically distinguishable; and, moreover, any difference between them has no material impact on results presented herein.  Hence, for the sake of simplicity, we choose to define the DB pion PDA by Eq.\,\eqref{phiH}.}
The $n=2,4$ moments in Eq.\,\eqref{moments} sit comfortably with the bounds derived in Ref.\,\cite{Bakulev:2011rp} and the sizable magnitude of the $n=6$ moment indicates that information is incorporated within our approach that is not expressed in Ref.\,\cite{Bakulev:2011rp}.

It is worth remarking, too, that on $Q^2\gtrsim 10\,$GeV$^2$, the solid (black) curve in Fig.\,\ref{figDSEprediction} also matches the AdS/QCD model result in Ref.\,\cite{Brodsky:2011xx}.  

One may contrast our scheme for the prediction of $G(Q^2)$ with the class of approaches that choose instead to infer a form of $\varphi_\pi$ by requiring agreement with the BaBar data, \emph{e.g}.\ Ref.\,\cite{Agaev:2010aq}.  A point of comparison here is provided by Table~II therein, which lists Gegenbauer-$3/2$ moments associated with the PDAs judged viable by this criterion.  These moments are defined via
\begin{equation}
\label{projection}
a_j(\zeta) = \frac{2}{3}\ \frac{2\,j+3}{(j+2)\,(j+1)}\int_0^1 dx\, C_j^{(3/2)}(2\,u-1)\,\varphi_\pi(u;\zeta)\,,
\end{equation}
where $\{C_j^{(3/2)},j=1,\ldots,\infty\}$ are Gegenbauer polynomials of order $\alpha=3/2$.  Using the DSE results, one finds (at $\zeta_B=2\,$GeV):
\begin{equation}
\label{Gmoments}
\begin{array}{l|cccccc}
    & a_2 & a_4 & a_6 & a_8 & a_{10}& a_{12}\\\hline
{\rm RL} & 0.23 & 0.11\phantom{7} & 0.066 & 0.045 & 0.033 & 0.025\phantom{3}\\
{\rm DB} & 0.15 & 0.057 & 0.031 & 0.018 & 0.013 & 0.0093
\end{array}\,.
\end{equation}
The root-mean-square relative-difference between these moments and those determined in Ref.\,\cite{Agaev:2010aq} is roughly 100\%.  Thus the PDA needed to reproduce the BaBar data is irreconcilable with that determined in \emph{ab initio} computations that unify the electromagnetic form factors of the charged and neutral pions.


\section{Conclusions and prospects}
We completed a computation of the $\gamma^\ast \gamma \to \pi^0$ transition form factor, $G(Q^2)$, in which all elements are determined by solutions of QCD's Dyson-Schwinger equations (DSEs) obtained in the rainbow-ladder (RL) truncation, the leading order in a systematic and symmetry-preserving approximation scheme.  In doing so we unified the description and explanation of this transition with the charged pion electromagnetic form factor, its valence-quark distribution amplitude, and numerous other quantities \cite{Roberts:2000aa, Chang:2011vu, Bashir:2012fs, Cloet:2013jya}, using a single DSE interaction kernel. 

The novel analysis techniques we employed made it possible to compute $G(Q^2)$ on the entire domain of spacelike momenta for the first time in a framework with a direct connection to QCD.  This enabled us to demonstrate that a fully self-contained and consistent treatment can readily connect a pion PDA that is a broad, concave function at the hadronic scale with the perturbative QCD prediction for the transition form factor in the hard photon limit.  Our prediction for $G(Q^2)$ agrees with all available data, except that obtained by the BaBar collaboration, and is fully consistent with the hard scattering limit.

It is worth emphasising that the normalisation of the $\gamma^\ast \gamma \to \pi^0$  transition form factor's hard scattering limit is set by the pion's leptonic decay constant, whose magnitude is fixed by the scale of dynamical chiral symmetry breaking, a crucial feature of the Standard Model.  Therefore, in order to claim an understanding of the Standard Model, it is critical to obtain new, accurate and precise transition form factor data on $Q^2>10\,$GeV$^2$ so that predictions such as that herein can reliably be tested.

\smallskip

\section*{Acknowledgments}
We are grateful for astute remarks by I.\,C.~Clo\"et, B.~El-Bennich, S.-X.~Qin, J.~Rodriguez Quintero and A.\,W.~Thomas; and for participation in the following workshops, which facilitated this research:
\emph{Many Manifestations of Nonperturbative QCD under the Southern Cross}, Ubatuba, Brazil (LC, AB, CDR, PCT);
\emph{2$^{\it nd}$ Workshop on Perspectives in Nonperturbative QCD}, IFT-UNESP, S\~ao Paulo, Brazil (AB, CDR, PCT);
\emph{Connecting Nuclear Physics and Elementary Particle Interactions: Building Bridges at the Spanish Frontier}, Punta Umbr\'{\i}a, Spain (KR, AB, CDR, PCT)
and
\emph{5$^{\it th}$ Workshop on Nonperturbative Aspects of Field Theories}, Morelia, M\'exico (KR, AB, JC-M, LXG, CDR, PCT).
Research supported by:
CIC (UMSNH) and CONACyT Grant nos.\ 4.10 and CB-2014-22117;
 U.S.\ Department of Energy, Office of Science, Office of Nuclear Physics, under contract no.~DE-AC02-06CH11357;
 and U.S.\ National Science Foundation, grant no.\ NSF-PHY1206187.

\appendix
\setcounter{figure}{0}
\setcounter{table}{0}
\renewcommand{\thefigure}{\Alph{section}.\arabic{figure}}
\renewcommand{\thetable}{\Alph{section}.\arabic{table}}

\section{Perturbation Theory Integral Representations}
\label{appA}
Here we describe the interpolations used in our evaluation of the transition matrix element in Eq.\,\eqref{anomalytriangle}.  The dressed-quark propagator is represented as \cite{Bhagwat:2002tx}
\begin{equation}
S(p) = \sum_{j=1}^{j_m}\bigg[ \frac{z_j}{i \gamma\cdot p + m_j}+\frac{z_j^\ast}{i \gamma \cdot p + m_j^\ast}\bigg], \label{Spfit}
\end{equation}
with $\Im m_j \neq 0$ $\forall j$, so that $\sigma_{V,S}$ are meromorphic functions with no poles on the real $p^2$-axis, a feature consistent with confinement \cite{Bashir:2012fs}.  We find that $j_m=2$ is adequate.

With relative momentum defined via $\eta=1/2$, we represent the scalar functions in Eq.\,\eqref{Gammapi} $({\cal F}=E,F,G)$ as a sum of two terms:
\begin{equation}
\label{Gpifit}
{\cal F}(k;P) = {\cal F}^{\rm i}(k;P) + {\cal F}^{\rm u}(k;P) \,,
\end{equation}
where that describing the infrared behaviour, labelled ``i'', is expressed via the following perturbation theory integral representation:
\begin{eqnarray}
\nonumber {\cal F}^{\rm i}(k;P) & = & c_{\cal F}^{\rm i}\int_{-1}^1 \! dz \, \rho_{\nu^{\rm i}_{\cal F}}(z) \bigg[
a_{\cal F} \hat\Delta_{\Lambda^{\rm i}_{{\cal F}}}^4(k_z^2) \\
&& \rule{7em}{0ex}
+ a^-_{\cal F} \hat\Delta_{\Lambda^{\rm i}_{\cal F}}^5(k_z^2)
\bigg], \label{Fifit}
\end{eqnarray}
and the ultraviolet ``u'' term is expressed analogously:
{\allowdisplaybreaks
\begin{subequations}
\begin{eqnarray}
\label{EFit}
E^{\rm u}(k;P) & = & c_{E}^{\rm u} \int_{-1}^1 \! dz \, \rho_{\nu^{\rm u}_E}(z)\,
 \hat \Delta^{1+\mathpzc{g}}_{\Lambda^{\rm u}_{E}}(k_z^2)\,,\\
F^{\rm u}(k;P) & = & c_{F}^{\rm u} \int_{-1}^1 \! dz \, \rho_{\nu^{\rm u}_F}(z)\,
 \Lambda_F^{\rm u} k^2 \Delta_{\Lambda^{\rm u}_{F}}^{2+\mathpzc{g}}(k_z^2)\,,\\
G^{\rm u}(k;P) & = & c_{G}^{\rm u} \int_{-1}^1 \! dz \, \rho_{\nu^{\rm u}_G}(z)\,
 \Lambda_G^{\rm u}\Delta_{\Lambda^{\rm u}_{G}}^{2+\mathpzc{g}}(k_z^2)\,, \label{Gufit}
\end{eqnarray}
\end{subequations}}
\hspace*{-0.5\parindent}with $\rho_\nu(z)$ defined in Eq.\,\eqref{rhonu}, $\hat \Delta_\Lambda(s) = \Lambda^2 \Delta_\Lambda(s)$, $k_z^2=k^2+z k\cdot P$, $a^-_E = 1 - a_E$, $a^-_F = 1/\Lambda_F^{\rm i} - a_F$, $a^-_G = 1/[\Lambda_G^{\rm i}]^3 - a_G$.  As elsewhere \cite{Chang:2013pqS, Chang:2013niaS}, $H(k;P)$ is small, has little impact, and is thus neglected.

The strength of the interaction detailed in Ref.\,\cite{Qin:2011dd} is specified by a product: $D\omega = m_G^3$.  With $m_G$ fixed, results for properties of ground-state vector and flavour-nonsinglet pseudoscalar mesons are independent of the value of $\omega \in [0.4,0.6]\,$GeV.  Using $\omega =0.5\,$GeV, the RL kernel yields $f_\pi=0.092\,$GeV with $m_G^{\rm RL}(\zeta=2\,GeV)=0.87\,$GeV.  The values of the interpolation parameters, which reproduce the associated numerical solutions of the gap and Bethe-Salpeter equations when used in Eqs.\,\eqref{Spfit}, \eqref{Gpifit}, are listed in Table~\ref{Table:parameters}.

\begin{table}[t]
\caption{Representation parameters. \emph{Upper panel}: Eq.\,\protect\eqref{Spfit} -- the pair $(x,y)$ represents the complex number $x+ i y$.  \emph{Lower panel}: Eqs.\,\protect\eqref{Fifit}--\protect\eqref{Gufit}.  (Dimensioned quantities in GeV).
\label{Table:parameters}
}
\begin{center}
%

\begin{tabular*}
{\hsize}
{
c@{\extracolsep{0ptplus1fil}}
c@{\extracolsep{0ptplus1fil}}
c@{\extracolsep{0ptplus1fil}}
c@{\extracolsep{0ptplus1fil}}
c@{\extracolsep{0ptplus1fil}}}\hline
$z_1$ & $m_1$  & $z_s$ & $m_2$ \\

$(0.44,0.014)$ & $(0.54,0.23)$ & $(0.19,0)$ & $(-1.21,-0.65)$ \\
\end{tabular*}

\begin{tabular*}
{\hsize}
{
l@{\extracolsep{0ptplus1fil}}
c@{\extracolsep{0ptplus1fil}}
c@{\extracolsep{0ptplus1fil}}
c@{\extracolsep{0ptplus1fil}}
c@{\extracolsep{0ptplus1fil}}
c@{\extracolsep{0ptplus1fil}}
c@{\extracolsep{0ptplus1fil}}
c@{\extracolsep{0ptplus1fil}}
c@{\extracolsep{0ptplus1fil}}}\hline
    & $c^{\rm i}$ & $c^{u}$ & $\phantom{-}\nu^{\rm i}$ & $\nu^{\rm u}$ & $a$\phantom{00} & $\Lambda^{\rm i}$ & $\Lambda^{\rm u}$\\\hline
E & $1 - c^{u}_E$ & $0.03$ & $-0.74$ & 1.08
    & 2.75\phantom{$/[\Lambda^{\rm i}_G]^3$} & 1.32 & 1.0\\
    F & 0.51 & $c^{\rm u}_E/10$ & $\phantom{-}0.96$ & 0.0
    & 2.78$/\Lambda^{\rm i}_{F}$\phantom{00} & 1.09 & 1.0 \\
G & $0.18$ & 2$\,c^{\rm u}_F$ & $\phantom{-}\nu^{\rm i}_F$ & 0.0 & 5.73$/[\Lambda^{\rm i}_G]^3$ & 0.94 & 1.0 \\\hline
\end{tabular*}
\end{center}

\vspace*{-4ex}

\end{table}

We remark that in checking the accuracy of the interpolations in Ref.\,\cite{Chang:2013pqS}, we found that a 4\% increase in $\nu_E^{\rm i}$ and introduction of ${\mathpzc g}=0.085$ in Eq.\,\eqref{EFit} provides a marginal improvement.  The latter is connected with an aspect of the generalised spectral representations that was explained in Ref.\,\cite{Chang:2013niaS}.  Namely, DSE kernels that preserve the one-loop renormalisation group behaviour of QCD will necessarily generate propagators and Bethe-Salpeter amplitudes with a nonzero anomalous dimension $\gamma_F$, where $F$ labels the object concerned.  Consequently, the spectral representation must be capable of describing functions of $\mathpzc{s}=p^2/\Lambda_{\rm QCD}^2$ that exhibit $\ln^{-\gamma_F}[\mathpzc{s}]$ behaviour for $\mathpzc{s}\gg 1$.  This is readily achieved by noting that
\begin{equation}
\ln^{-\gamma_F} [D(\mathpzc{s})]
= \frac{1}{\Gamma(\gamma_F)} \int_0^\infty \! dx\, x^{\gamma_F-1}
\frac{1}{[D(\mathpzc{s})]^x}\,,
\end{equation}
where $D(\mathpzc{s})$ is some function.  Such a factor can be multiplied into any existing spectral representation in order to achieve the required ultraviolet behaviour.  Importantly, for practical applications involving convergent four momentum integrals, like those generated by Eq.\,\eqref{anomalytriangle}, it is adequate to develop and use a power law approximation; viz.,
\begin{equation}
\ln^{\gamma_F} [D(\mathpzc{s})] \approx [D(\mathpzc{s})]^{\mathpzc g}.
\end{equation}
With ${\mathpzc g}$ chosen appropriately, this is accurate on the material domain and greatly simplifies the subsequent numerical calculation.  Owing to the quark-level Goldberger-Treiman relation \cite{Maris:1997hd, Maris:1997tm, Qin:2014vya}:
\begin{equation}
f_\pi^0 E_{\pi}^0(k^2,k\cdot P=0;P^2=0) =B_0(k^2) \,, \label{BGTrelation}
\end{equation}
where the superscript ``0'' indicates a quantity computed in the chiral-limit, the same factor ${\mathpzc g}$ must be used to amend the power-law behaviour of $\sigma_S(p^2)$ in Eq.\,\eqref{GenSp} via Eq.\,\eqref{Spfit}.


\end{document}